\documentclass[12pt]{article}
\usepackage{graphicx}
\usepackage{latexsym}
\usepackage[left=.75in,top=1in,right=.75in,bottom=1in]{geometry}
\usepackage{amsmath}

    \def\independenT#1#2{\mathrel{\setbox0\hbox{$#1#2$}%
    \copy0\kern-\wd0\mkern4mu\box0}} 

\newenvironment{packed_item}{
\begin{itemize}
 \setlength{\itemsep}{0pt}
  \setlength{\parskip}{0pt}
  \setlength{\parsep}{0pt}
}{\end{itemize}}

\usepackage{amssymb}

\usepackage{multirow}
\usepackage[numbers, sort&compress]{natbib}
\usepackage[section]{placeins}

\usepackage[font={footnotesize,it}]{caption}

\usepackage{url}

\usepackage{pdflscape}
\usepackage{arydshln}

\usepackage[dvipsnames]{xcolor}

\usepackage[normalem]{ulem} 
\usepackage[makeroom]{cancel}

\title{Brief Article}
\author{The Author}
\date{November 10, 2021}                                           % Activate to display a given date or no date

\begin{document}

\begin{centering}
\Large{Sustainable East Africa Research in Community Health \\ (SEARCH) Collaboration

\vspace{3em}

{\textbf{ Statistical Analysis Plan for Health Outcomes \\
in Phase 1 of the SEARCH-IPT Study}}
\vspace{4em}

 Laura B. Balzer, PhD$^1$ \\
Joshua Nugent, MS$^1$\\
Diane V. Havlir, MD$^2$ \\
Gabriel Chamie, MD$^2$ 
\vspace{4em}

November 10, 2021 \\
Version 2.1}

\end{centering}
\vfill
$^1$University of Massachusetts, Amherst; $^2$University of California, San Francisco

\newpage

\tableofcontents 

% CHECK MOH and SEARCH-IPT

\section{Brief study overview}

The SEARCH-IPT Study is a cluster randomized trial designed to evaluate whether a multicomponent intervention increases uptake of isoniazid (INH) preventive therapy (IPT) and reduces the incidence of tuberculosis (TB) in Uganda (Clinicaltrials.gov: NCT03315962). Details of the multiphase trial design and procedures can be found in the Study Protocol. Here, we provide the statistical analysis plan for health outcomes in Phase 1 of the study. Analysis plans for qualitative outcomes, cost-effectiveness outcomes, and business outcomes are available elsewhere. A history of changes to this analysis plan is given in the Appendix.

\textbf{Study design:} Phase 1 of the SEARCH-IPT Study aims to increase provider prescription and patient use of IPT among adults in HIV care in Uganda. Within the Eastern, East-Central, and Southwestern regions of Uganda, districts were first grouped into 14 clusters of 4-7 districts. To improve statistical power \citep{HayesMoulton2009,Balzer2015Adaptive}, these groups were pair-matched on characteristics expected to be predictive of IPT uptake: region, number of adults in active HIV care, urbanicity, and presence of a SEARCH universal HIV “test and treat” trial site. Within each pair, the groups were then randomized to intervention or the standard-of-care. There are 7 groups in each arm. Formal power calculations are given in the Appendix.  

\textbf{Study arms:} In districts randomized to the intervention, the groups are the mini-collaboratives in which the intervention package is delivered to District Health Officers (DHOs) and District TB/Leprosy Supervisors (DTLSs), mid-level health managers who oversee health service delivery (DHOs) and TB-specific activities (DTLSs) in Uganda. The SEARCH-IPT intervention package is detailed in the Study Protocol. In brief, the social network-based, behavioral change model consists of teaching mini-collaboratives with business leadership and management training; reporting, review and and discussion of IPT uptake at mini-collaborative meetings; and a two-way SMS system for DHOs/DTLSs and their front-line providers to help enable IPT uptake. 

Districts randomized to the control receive no study interventions, beyond ensuring that the Ministry of Health  IPT guidelines are disseminated. 

\textbf{Study timeline:} As described elsewhere, there were many events unrelated to the study, which are likely to impact health outcomes in Phase 1. Immediately prior to study launch (end of 2017), there was a massive shortage of IPT that lasted until the end of 2018. Then starting in the third quarter of 2019 (Q3-2019), the Ministry of Health led a country-wide program to increase IPT use with support from PEPFAR (hereafter called “the 100-day IPT push”). Additionally in 2019, the Ugandan government divided select districts into two new districts. Finally, the COVID-19 pandemic began in March 2020. 

To account for drug shortages and answer the most relevant scientific question, the measurement period for the main study outcomes (IPT uptake and HIV-associated TB incidence) will begin at the quarter corresponding to the first-year meeting of the mini-collaboratives, which occurred in the intervention arm only. Specifically, follow-up begins in Q1-2019 in Southwestern districts and in Q2-2019 in the Eastern and East-Central districts. All districts are followed through the close of Phase 1, and analyses will be over two years (8 quarters) of follow-up. Data from the quarter immediately preceding the follow-up period (i.e., Q4-2018 in the Southwest and Q1-2019 in the East and East-Central) are assumed to provide baseline data. 

\textbf{Descriptive analyses:} At baseline, we will describe the study population by region (East, East-Central, and Southwest) by providing summary measures of district-level characteristics, including but not limited to \vspace{-1em}
\begin{packed_item}
\item Number of districts 
\item Number of participating DHOs/DTLSs
\item Number of adults in active HIV care at the district-level 
\item Number of adults in active HIV care in two largest clinics in each district 
\item HIV prevalence
\item Incidence of IPT uptake 
\item Incidence of HIV-associated TB disease
\end{packed_item}
We will also provide these summary measures by arm. We will additionally provide descriptions of secular events and non-study activities that are expected to impact study outcomes. 

\section{Primary outcome: IPT uptake}

The primary endpoint is ``IPT uptake'': the rate at which eligible HIV-positive adults receive an IPT prescription in health facilities overseen by DHOs participating in the SEARCH-IPT trial. This endpoint will be measured over 8 quarters (starting Q1-2019 in the Southwest and in Q2-2019 in the East and East-Central) among adults in active HIV care in two clinics in each district in the trial. These clinics were selected based on size, generally the biggest, and are assumed to be representative of the overall district. (This assumption of generalizability may or may not be reasonable in smaller and more rural districts.)

\textbf{Data source:} For each \emph{clinic}, aggregate data on the total number of IPT starts and the total number of adults in active HIV care \emph{in each quarter} will be extracted from the Ministry of Health HMIS database. Data on IPT initiation are available by sex, and data on adults in active care are available by sex and 1st/2nd/3rd line ARV regime. Primary analyses will pool over sex and ARV regime. Secondary analyses will be stratified on sex. 

\textbf{Missing data:} There is no reason to believe data will be differentially captured by arm. Additionally, we are unable to differentiate between missing data on IPT prescriptions and simply 0 prescriptions for that clinic-quarter. Therefore, if data are missing on IPT prescriptions for a given clinic-quarter, we will assume there were 0 starts in that clinic-quarter; this will provide a lower bound on the total IPT starts for that clinic. However, if for a given clinic, data on IPT prescriptions are missing for all quarters, we will exclude that clinic.

Likewise, there is no way to differentiate between missing data on the number of patients 2nd/3rd line ARV regimens or simply 0 participants on those lines. If data on the number of patients on 2nd or 3rd line ARV regimen are missing, we will assume that 0 patients are on those lines for that clinic-quarter. We do not anticipate data on the active care size to be completely missing for a given clinic-quarter; if this happens, we will estimate the clinic-quarter active care size by averaging the size in the prior clinic-quarter with the size in the subsequent clinic-quarter. 

\textbf{Incidence rate calculations:} Because the number of adults in HIV care changes over time and the probability of starting IPT is not uniform over the 8 quarters of follow-up, we focus on estimation and inference of the incidence rate of IPT initiation, as opposed to the cumulative incidence. For each clinic, we will determine the total number of IPT starts and the total person-time-at-risk of IPT initiation (described below) over the two-year follow-up period. In primary analyses, we will calculate the district-level incidence rate as the total starts in a district (summed over the 2 clinics) divided by the total person-time-at-risk for that district (again summed over the 2 clinics). In secondary analyses, we will calculate the district-level incidence rate by averaging the clinic-specific incidence rates within each district. 

Because data for each clinic-quarter are only available at the aggregate-level, we will take the following approach to calculating the person-time-at-risk (PT). In line with standard practice \citep{RothmanModern}, we will assume that IPT initiations occur uniformly in each clinic-quarter. In primary analyses, we will also assume that all who started IPT during follow-up remain in active HIV care for the remainder of follow-up. While the latter assumption is supported by prescribing practices (i.e., to patients who are stable in care), we will conduct two sensitivity analyses. In the first, we will assume that 97.5\% of persons who previously started IPT remain in active care at the same clinic from quarter-to-quarter. In the second, we will assume that 0\% of persons who previously started IPT remain in active care from quarter-to-quarter. The second sensitivity analysis is expected to be extremely conservative and will provide a lower bound in the estimated incidence rates. Altogether, PT for each clinic-quarter will be calculated as the quarter-specific number in active HIV care minus half the IPT starts in that quarter and X\% of the previous IPT starts, where X\%=100\% in the primary analysis, 97.5\% in the first sensitivity analysis, and 0\% in the second sensitivity analysis. The resulting incidences rates will be converted from being in `clinic-quarters' to `person-years' for ease of interpretation.

In a secondary analysis, we will calculate the two-year, cumulative incidence of IPT initiation as the total number of IPT starts, divided by the average active care size over follow-up. As before, we will calculate the district-level cumulative incidence in two ways: (1) the total starts in a district (summed over the 2 clinics) divided by the total active care size for that district (again summed over the 2 clinics); and (2) by averaging the clinic-specific cumulative incidences within each district. We will also report total starts by arm. 

All analyses will assume that all adults in active HIV care are eligible for IPT at the beginning of follow-up (Q1-2019 in the Southwest and Q2-2019 in the East/East-Central). By randomization, the number in active HIV care and the number who had previously started IPT should be balanced between arms.

\textbf{Effect estimation:} To estimate the effect of the SEARCH-IPT intervention on IPT uptake, we will compare IPT incidence rates by arm using a two-stage approach that appropriately accounts for the cluster randomized study design \citep{HayesMoulton2009}. In Stage 1, the incidence of IPT uptake will be estimated for each district, as described above. Then in Stage 2, uptake will be compared between arms with targeted minimum loss-based estimation (TMLE) \citep{MarkBook}, adaptively selecting the district-level adjustment variables that optimize precision, while preserving Type-I error control \citep{Balzer2016DataAdapt,Balzer2021twostage}. Specifically, we will use leave-one-out cross-validation to select from the following set of baseline covariates the ones which maximize empirical efficiency: baseline IPT uptake, baseline active care size, or nothing (unadjusted). In secondary analyses, we will implement an unadjusted effect estimator as the contrast in the average IPT uptake between arms. 

The primary effect measure will be on the relative scale and for the $N$ districts involved in the study (i.e., the sample rate ratio) \citep{Neyman1923,Rubin1990,Imbens2004, Imai2008, Balzer2016SATE}: 
\begin{align*}
\frac{\psi(1)}{\psi(0)} = \frac{1/N \sum_{i=1}^N \alpha_i Y_i(1) }{1/N \sum_{i=1}^N \alpha_i Y_i(0)}
\end{align*}

where $\alpha_i$ are weights for the district $i$ and $Y_i (z)$ is the counterfactual IPT uptake over the two-years of follow-up if possibly contrary-to-fact district $i$ were in treatment arm $Z=z$. We will test the \emph{null hypothesis} that the SEARCH-IPT intervention did not improve IPT uptake, as compared to standard-of-care, with a one-sided test at the 5\% significance level. In secondary analyses, we will examine the effect on the absolute scale: $\psi(1)-\psi(0)$. We will report point estimates and two-sided 95\% confidence intervals for the arm-specific endpoints as well as the relative and absolute effects. 

Primary analyses will weight the $N$ districts equally: $\alpha_i$=1 for all $i$. In secondary analyses, we will evaluate the effect at the clinic-level, and in sensitivity analyses, we will weight proportionally to active care size \citep{Benitez2021}.  

Primary analyses will include all districts, including those created during follow-up. Secondary analyses will restrict to the original districts, who had the longest exposure to the intervention.  

\textbf{Statistical inference:} Recall that 4-7 districts were put into groups for the purposes of pair-matching and treatment randomization. In the intervention arm, these clusters are the basis for mini-collaborative activities. In the control arm, there are \textbf{no} mini-collaborative meetings or any other study activities. Instead, the control condition corresponds with the standard of care. Therefore, it is reasonable to assume districts in the control arm are independent. 

Given this independence structure (at the mini-collaborative level in the intervention arm and at the district level in the control arm), we take the following approach to statistical inference. In the primary analysis, we will assume there are 7 independent units corresponding to the mini-collaboratives in the intervention arm, and 40 independent units corresponding to the districts in the control arm (total $N=47$). In sensitivity analyses, we will treat the randomized cluster as the independent unit (total $N=14$). In all analyses, we will use the influence function for standard error estimation \citep{Balzer2016SATE} and the Student’s $t$-distribution with $(N-2)$ degrees of freedom as finite sample approximation to the standard normal distribution \citep{HayesMoulton2009}.

Primary analyses will break the matched pairs used for randomization. For completeness, we will also conduct a paired t-test ($N=7$/arm), implement a permutation test, and also estimate the intervention effect with mixed (random) effects models and with generalized estimating equations (GEE) \citep{LairdWare82,LiangZeger86}.

\textbf{Sensitivity analyses}: In addition to the sensitivity analyses described above, we will conduct the following sensitivity analyses. To minimize the impact of secular trends, we will exclude the ``100-day IPT push'' (Q3-2019). To better understand the impact the COVID-19 pandemic which caused a lockdown to be initiated in Q2-2020, we will  conduct an analysis excluding the post-lockdown period and  an analysis excluding the pre-lockdown period. Given the expected skewed distribution of clinic sizes, we will conduct two sensitivity analyses excluding clinics with baseline active care size exceeding 2500 patients and exceeding 5000 patients. 

\textbf{Subgroup analyses:} We will repeat the above analyses within strata defined by sex, baseline active care size ($\leq$ 1000 vs. $>$1000), and region. 

\textbf{Additional analyses:} We will examine how IPT uptake varies over time by plotting the quarterly incidence rate (as described above) and the cumulative incidence (quarter-specific starts divided by quarter-specific number of adults on ART) over time by arm, region, and sex. We will also formally evaluate trends over time using TMLE to estimate parameters of a marginal structural model and using GEE \citep{MarkBook, LiangZeger86, Petersen2014ltmle}. We will also examine how DHO/DTLSs participation mediates the intervention effect with TMLE. Finally, we will conduct an “as-treated” analysis where intervention districts whose DHOs/DTLSs had poor participation in mini-collaboratives are analyzed as receiving the standard-of-care (i.e., control). 

\section{Secondary outcome: HIV-associated TB incidence}

We will also examine the impact of the SEARCH-IPT intervention on ``HIV-associated TB incidence'': the rate at which persons in active HIV care are diagnosed with TB disease in districts overseen by DHOs participating in the SEARCH-IPT trial. Thus, the target and analytic population include all persons in active HIV care, regardless of their age, sex, ARV regimen, IPT eligibility, or IPT uptake. As with the primary endpoint, TB incidence will be measured over 8 quarters (starting Q1-2019 in the Southwest and in Q2-2019 in the East and East-Central).

\textbf{Data source:} For each \emph{district}, aggregate data on the total number of TB diagnoses among persons in active HIV care and the total number of persons in active HIV care \emph{in each quarter} will be extracted from the Ministry of Health HMIS database. While data on the active size are available by 1st/2nd/3rd line ARV regime, all analyses will pool ARV regime.

\textbf{Missing data:} There is no reason to believe data will be differentially captured by arm. As before, we are unable to distinguish between missing values and 0 persons diagnosed with TB. Therefore, if data are missing on TB cases for a given district-quarter, we will assume there were 0 cases in that district-quarter. However, if for a given district, data on TB cases are missing for all quarters, we will exclude that district from the analysis.

If data on the number of patients on 2nd or 3rd line ARV regimen are missing, we will again assume that 0 patients are on those lines in that district-quarter. We do not anticipate data on the active care size to be completely missing for a given district-quarter; if this happens, we will estimate the district-quarter active care size by averaging the size in the prior district-quarter with the size in the subsequent district-quarter. 

\textbf{Incidence calculations:} Because the number of persons in active HIV care changes over time and the rate of TB disease diagnosis may not be uniform over the 8 quarters of follow-up, we again focus on the incidence rate, as opposed to the cumulative incidence. For each district, we will estimate the TB incidence rate as the total number of cases divided by the total person-time-at-risk (PT) over the two-year follow-up period. As before, when calculating the incidence rate, we will assume events (here, TB disease diagnoses) occur uniformly in each district-quarter and that a fixed proportion of prior TB cases remain in active care within the same district from quarter-to-quarter. Then in each district-quarter, the PT will be calculated as the quarter-specific number in active HIV care minus half the cases occurring in that quarter and X\% of the previous cases, where X\%=100\% in the primary analysis, 97.5\% in the first sensitivity analysis, and 0\% in the second sensitivity analysis. The resulting incidences rates will again be converted from being in `clinic-quarters' to `person-years' for ease of interpretation. In secondary analyses, we will estimate the district-specific cumulative incidence of TB as the total number of cases divided by the average active care size. We will also report total number of TB diseases diagnoses by arm. 

All analyses will assume all persons in active HIV care are at risk of TB disease at the beginning of follow-up (Q1-2019 in the Southwest and Q2-2019 in the East/East-Central). By randomization, the number in active HIV care and the number who had previously been diagnosed with TB disease should be balanced between arms.

\textbf{Effect estimation \& statistical inference:} To estimate and obtain inference for the effect of the SEARCH-IPT intervention on HIV-associated TB, we will compare the district-level TB incidence rates by arms using an analogous two-stage approach as for the primary outcome. In Stage 1, the incidence of TB diagnoses will be estimated for each district, as described above. Then in Stage 2, district-level incidence will be compared between arms with TMLE to adaptively select from the following set of district-level adjustment variables the ones that optimize precision: baseline TB incidence, baseline active care size, or nothing (unadjusted). In secondary analyses, we will implement the unadjusted effect estimator as the contrast in the average TB incidence rates between arms. We will test the \emph{null hypothesis} that the SEARCH-IPT intervention did not reduce the incidence of HIV-associated TB, as compared to standard-of-care, with a one-sided test at the 5\% significance level. 

As before, the primary effect measure will be on the relative scale and for the $N$ districts involved in the study (i.e., the sample incidence ratio). In secondary analyses, we will examine the effect on the absolute scale. We will also report point estimates and two-sided 95\% confidence intervals for the arm-specific endpoints and effect estimates. Primary analyses will weight the $N$ districts equally, while sensitivity analyses will weight proportionally to baseline active care size \citep{Benitez2021}. We will use the same approach to statistical inference as the primary endpoint. 

\textbf{Sensitivity analyses:} We will also conduct the sensitivity analyses, as specified for the primary endpoint. Here, we will exclude districts whose baseline active care size is $>$10,000 patients and $>$20,000 patients. 

\textbf{Subgroup analyses:} We will repeat the above analyses within strata defined by baseline active care size ($\leq$ 5000 vs. $>$5000), region, and age (pending data availability). 

\textbf{Additional analyses:} We will examine how TB incidence varies over time by plotting the quarterly incidence rate (as described above) and the cumulative incidence (quarter-specific number of TB cases divided by quarter-specific number on ART) over time by arm and region. Using analogous methods to the primary endpoint, we will evaluate time trends and the impact of DHO/DTLSs participation.  

\textbf{Secondary endpoint:} Pending data availability,  we will repeat the above analyses to examine the SEARCH-IPT effect on TB treatment initiation.%, calculated as for TB incidence.

\section{Secondary outcome: Knowledge, attitudes, and practices regarding IPT}

To assess mechanisms through which the intervention operates, we will conduct annual quantitative surveys among DHOs and DTLSs to evaluate changes in their knowledge, attitudes, and practices regarding IPT. Details of these surveys are available elsewhere; here, we focus on the statistical analysis to evaluate change from baseline in familiarity with IPT, knowledge of IPT’s health benefits, and practical challenges in TB management. Specific survey questions and their coding are given in the Appendix.

\textbf{Data source:} Surveys will be administrated to all DHOs and DTLSs by study staff at the time of randomization and then 1 and 2 years later. Study staff will attempt to reach and survey all participants in both arms. 

As an example of a question, consider \emph{``On a scale of 1 to 5, with 5 being very familiar with IPT, and 1, not knowing much yet about IPT, how familiar are you with IPT?''} The possible answers are \emph{``1: No knowledge of IPT; 2: Somewhat unfamiliar–low knowledge of IPT; 3: Somewhat familiar; 4: Familiar; 5: Very familiar–high knowledge of IPT''}. Thus, the survey questions will elicit Likert-type answers with evenly spaced responses. In line with standard practice for such outcomes \citep{HayesMoulton2009}, we will evaluate them quantitatively. Specifically, for each district, we will calculate the average response among its representatives at baseline, at year-1, and at year-2. 

\textbf{Effect estimation:} To estimate and obtain inference for the effect of the SEARCH-IPT intervention on IPT knowledge/attitudes/practices, we will first calculate a change-outcome as the average response at follow-up minus the average response at baseline in each district. We will then compare these outcomes with the unadjusted effect estimator (equivalent to the Student’s $t$-test) and test the \emph{null hypothesis} of no improvement in IPT knowledge/attitudes/practice, as compared to standard-of-care, with a one-sided test at the 5\% significance level. 

We will focus on effect measures on the absolute scale and for the $N$ districts with both baseline and follow-up data. Primary analyses will focus on the change from baseline to year-1. Secondary analyses will focus on change from baseline to year-2 and from year-1 to year-2. We will report point estimates and two-sided 95\% confidence intervals for the arm-specific annual responses, the arm-specific estimates of change, and the effect estimates. Primary analyses will weigh the $N$ districts equally and include all districts, including those created during follow-up. We will use the same approach to statistical inference as the primary endpoint. 

\textbf{Additional analyses:} We will repeat the above analyses stratifying on whether the respondent is a DHO or a DTLS. 

\section{Secondary outcome: IPT completion among persons starting IPT}

A key intermediate between IPT uptake and TB prevention is IPT completion. Therefore, we will also assess the SEARCH-IPT intervention on ``IPT completion'', defined as dispensation of a full course of IPT within 9 months of initiation. As detailed in the corresponding SOP, this endpoint will be measured among $N=800$ (400 persons/arm) who are aged 15+ years, HIV-infected, and started IPT at one of 16 facilities in Southwestern region. Within each clinic, these participants will be sampled such that half initiated IPT before Q3-2019 and half initiated in Q3-2019 (during the ``100-day IPT push'').

\textbf{Data source:} For each sampled participant, de-identified data on IPT completion (via INH pills dispensed over 9 months) will be obtained from the IPT register and blue cards. Additionally, data on the following variables will be recorded: gender, date of birth, marital status, ARV regimen, ARV switches while on IPT (date and new regimen), most recent HIV RNA viral load, if on Septrin or Dapsone while taking INH, and diagnosis of active TB. 

\textbf{Effect estimation:} When evaluating IPT completion among the subsample of initiators, we will evaluate the arm-specific and relative risk of IPT completion with an individual-level TMLE, adjusting for sex, age, and timing of IPT initiation. The details of this approach are given in van der Laan et al. \citep{vanderLaan2013Community}. In brief, this approach assumes the adjustment covariates are sufficient to ``block'' the impact of shared factors that may differ between clusters, here clinics. In other words, two participants from two different clinics would have the same outcome probability if they had the same gender, age, and IPT initiation date \emph{and} their respective clinics had received the same level of the treatment. This approach also relies on the assumption that sampled participants are conditionally independent, given the treatment assignment and the adjustment set. These assumptions are reasonable given that, beyond the intervention, the major driver of IPT prescribing practices and, thus, completion are secular factors (e.g., the Q3-2019 ``100-day IPT push'') and the participant’s timing of IPT initiation is included in the adjustment set. 

To flexibly control for sex, age, and timing of IPT initiation, we will implement TMLE with Super Learner, an ensemble machine learning algorithm that creates the best weighted combination of predictions from a set of candidate learners \citep{SuperLearner}. Our candidates will include multivariate adaptive regression splines, main terms logistic regression, and the simple mean. In sensitivity analyses, we will implement TMLE using main terms logistic regression instead of Super Learner. Additional sensitivity analyses will include marital status in the adjustment set and include penalized regression in the Super Learner library. 

We will also evaluate IPT completion using an analogous analysis as for the primary endpoint. Specifically, treating the clinic as the independent unit, we will use adaptive pre-specification to select among the following individual-level adjustment variables the ones that maximize precision in TMLE: sex, age, and timing of IPT initiation \citep{Balzer2016DataAdapt, Balzer2018Hierarchical}. We will also report unadjusted estimates from this approach as well as conduct an analysis weighting proportionally to the clinic’s total IPT starts and an analysis treating clinics as fixed effects. 

For all approaches, we will test the \emph{null hypothesis} of no improvement from the intervention with a one-sided test at the 5\% significance level. We will also report point estimates and two-sided 95\% confidence intervals for the relative effect, the absolute effect, and the arm-specific mean outcomes. We will also report total completions by arm.

\textbf{Additional analyses:} We will repeat the above analyses within subgroups defined by sex, timing of IPT initiation (before Q3-2019 ``100-day IPT push'' or during Q3-2019), age group (15-24 years or 25+ years), and marital status (never married, married/living together, divorced/separated/widowed, missing). We will also use an individual-level TMLE to examine predictors of non-completion. We will provide descriptive statistics on the subsample, including their demographics and care outcome (e.g., ARV switches). 

\textbf{Secondary endpoint:} We will repeat the above analyses to examine the effect of the SEARCH-IPT intervention on viral suppression, defined as HIV RNA viral level $<$400 copies/mL. 

\section{Appendix} 

\subsection{History of changes:}
\begin{packed_item}
\item Version 1.0 was locked on July 18, 2021 prior to unblinding and effect estimation for IPT uptake (the primary outcome), IPT completion among persons starting IPT, and HIV-associated TB incidence. 
\item Version 2.0 was created to include the pre-specified analysis plan for the quantitative surveys on knowledge, attitudes, and practices regarding IPT. Version 2.0 was locked on September 15, 2021, prior to unblinding and effect estimator for the survey outcomes. 
\item Version 2.1, this version, was created to correct grammatical errors and other typos. Version 2.1 was locked on November 10, 2021 prior to submission of the manuscript.
\end{packed_item}

\subsection{Select survey questions:}
The following are examples of survey questions, designed to understand the impact of the SEARCH-IPT intervention on IPT knowledge, attitudes, and practices among DHOs and DTLSs. Full survey materials are available elsewhere.
\begin{enumerate}
\item How familiar are you with IPT? With responses: 1: no knowledge of IPT; 2: somewhat unfamiliar – low knowledge of IPT; 3: somewhat familiar; 4: familiar; 5: very familiar – high knowledge of IPT.
\item	How strong is the evidence that INH prevents active TB in HIV-infected patients? With responses: 1: very weak; 2: weak; 3: mixed; 4:  strong; 5: very strong.
\item	How strong is the evidence that INH prevents death in HIV-infected patients? With responses: 1: very weak; 2: weak; 3: mixed; 4:  strong; 5: very strong.
\item	How much risk of INH drug resistance is there if a person develops active TB after completing IPT? With responses: 1: no risk of INH resistance; 2: low risk; 3: moderate risk; 4: high risk; 5: very high risk.
\item	How difficult is it for providers in this district to add INH to standard care for HIV-infected people in order to prevent TB? With responses: 1:  very easy; 2: easy; 3: neither hard nor easy; 4:  difficult; 5: very difficult.
\item	How hard is it to influence changes in practice among frontline providers around TB management? With responses: 1: very easy; 2: easy; 3: neither hard nor easy; 4: difficult; 5: very difficult.
\end{enumerate}

\subsection{Power calculations: }

Sample size and power calculations were based off the standard formulas for one-sided tests in cluster randomized trials with an incidence rate endpoint \citep{HayesMoulton2009}. As a conservative approximation, these calculations considered the randomization groups to be the independent unit. We also expect these calculations to be conservative, because of the precision gained through covariate adjustment in the analysis. 

We estimated 14 groups (7 groups/arm) would provide 80\% power to detect at least a 10\% absolute increase in IPT uptake (the primary outcome) from 22 per 100 person-years under the control, assuming a coefficient of variation of $k=0.25$ and $\approx$ 21,500 person-years of follow-up in each group. As shown in the following Figure, these calculations are fairly insensitive to expected amount of follow-up in group. If uptake is greater than expected (e.g., 30 per 100 person-years) under the control, we anticipate remaining well-powered to detect at least a 13.5\% increase in uptake.  

\begin{figure}[!h]
    \centering
\includegraphics[width=0.5\textwidth]{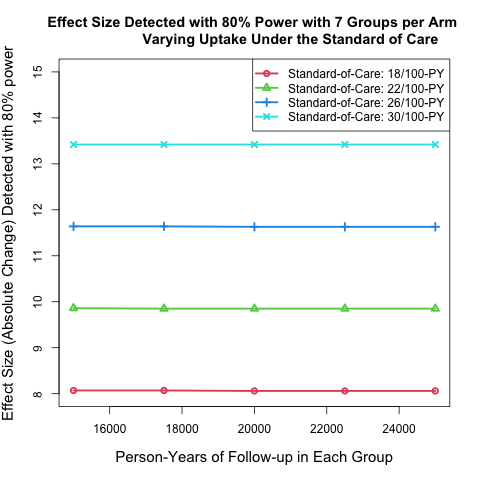}
    %\caption{Caption}
   % \label{fig:my_label}
\end{figure}

\bibliography{Bibliography}

\end{document}